\documentclass{PoS}

\title{Cooling of Color Superconducting Compact Stars}

\ShortTitle{Cooling of Compact Stars}

\author{{David Blaschke}\\ 
Theory Group, GSI Darmstadt, D-64291 Darmstadt, Germany\\
Bogoliubov Laboratory for Theoretical Physics, JINR Dubna, 141980 Dubna, 
Russia\\
E-mail: \email{blaschke@theory.gsi.de}}

\abstract{
We review the status of research on the cooling of compact stars, with
emphasis on the influence of color superconducting quark matter phases. 
Although a consistent microscopic approach is not yet available, 
severe constraints on the phase structure of matter at high densities come 
from recent mass and cooling observations of compact stars. 
}

\FullConference{Johns Hopkins Workshop on 'Strong Matter in the Heaven',
Budapest, August 1-3, 2005}


\begin{document}

\section{Introduction}
The fundamental questions for the origin of the mass of hadrons (chiral 
symmetry breaking), or the mechanism for confinement of quarks and gluons 
require answers from QCD as the fundamental quantum field theory of strong 
interactions in its non-perturbative domain.
Despite progress with lattice QCD simulations and systematic approaches
such as Dyson-Schwinger equations and renormalization-group equations, 
effective models for the strong interaction will remain indispensable for 
bridging the gap to phenomenological results accumulated in nuclear and 
particle physics as well as in astrophysics.  

A major challenge of experimental programs at large-scale 
facilities such as CERN Geneva, BNL Brookhaven or GSI Darmstadt is
to produce hadronic matter under extreme conditions of 
temperature and density in ultrarelativistic heavy-ion collisions 
in order to investigate the phase transition to a state of matter where 
the chiral symmetry and asymptotic freedom (deconfinement) of the QCD 
Lagrangian are restored. 
Quite opposite to the exploration of these phase transitions 
under terrestrial laboratory conditions, 
the strongly interacting matter in the heaven, namely in compact stars, 
is not plagued by the limitations of small volumes, short timescales and 
strong nonequilibrium and provides a unique view into the phase diagram
at low temperatures and high densities, see Fig. 1. 
\begin{figure}[h]
\includegraphics[width=0.8\textwidth,angle=0]{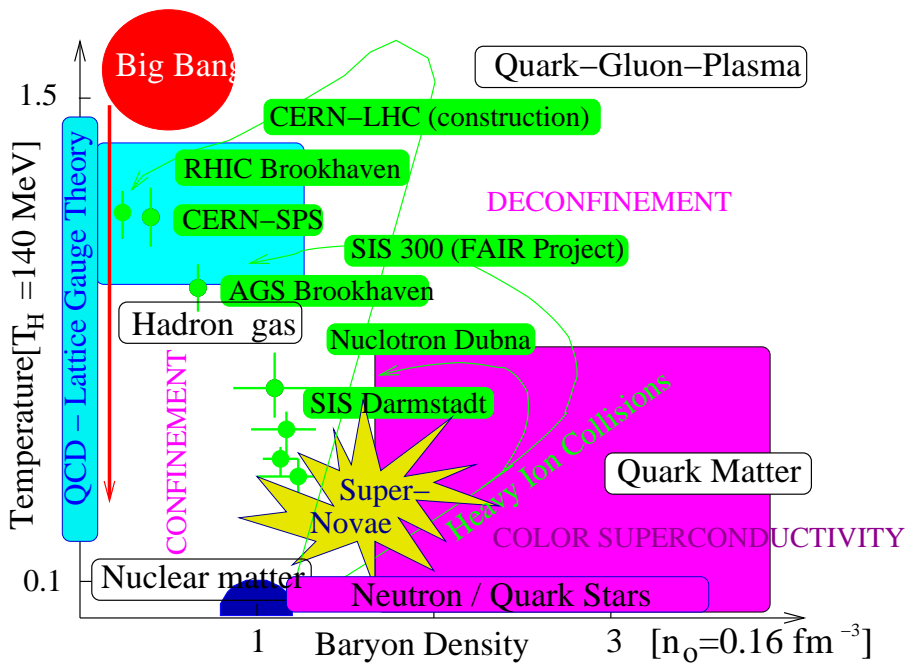}
\caption{The phase diagram fo strongly interacting matter is under exploration 
in Lattice gauge theory simulations, heavy-ion collisions and astrophysics of 
compact stars.}
\label{fig0a}
\end{figure}
In this region quark matter, being a cold and dense Fermi system with 
attractive interactions, is expected to appear in a color-superconducting 
state due to the instability against Cooper-pairing and formation of diquark  
Bose condensates.
The present contribution reviews recent investigations of the question 
whether observations of cooling compact stars can provide constraints on the
development of microscopic approaches to phase diagram, equation of state (EoS)
and transport properties of dense QCD matter, performed along the lines of the 
scheme given in Fig. 2. 
\begin{figure}[h]
\includegraphics[width=0.8\textwidth,angle=0]{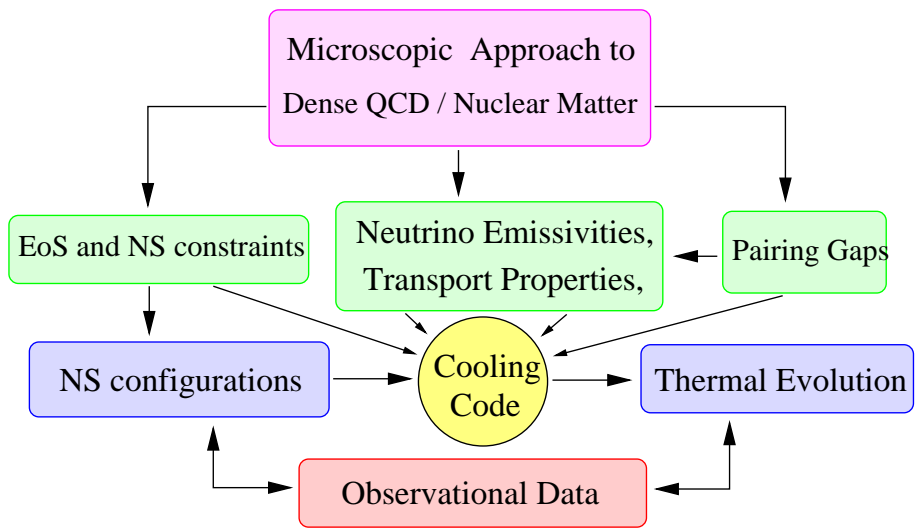}
\caption{Scheme for the interplay between different aspects of compact star 
phenomenology and microscopic theory of dense QCD matter. In the 
present contribution the focus is on aspects of the quark-hadron phase 
transition and effects of color superconductivity.}
\label{fig0b}
\end{figure}
Starting from a hadronic baseline for the EoS and cooling regulators we
introduce the new, stringent observational constraints on the masses, cooling 
curves and population of compact stars.
We present state-of-the-art phase diagrams of three-flavor QCD matter 
under compact star constraints for chiral quark models of the 
Nambu--Jona-Lasinio (NJL) type and examine possible hybrid star configurations
regarding the questions (i) do strange quark matter phases occur? and 
(ii) which patterns of color superconductivity are admissible?
 
\section{EoS and Structure of Hadronic Stars}
Hadronic matter in compact stars can reach densities above 1 fm$^{-3}$, so that
systematic investigations shall be based on relativistic quantum 
field-theoretical approaches as reviewed, e.g. in 
\cite{Glendenning:1997wn,Weber:1999qn}.  
Comparative studies often employ the representation of the energy per nucleon
\begin{equation}
E(n,\beta)=E_0(n) + \beta^2 E_s(n)~,
\end{equation}
where $n$ is the baryon number density, $\beta=1-2x$ the asymmetry parameter
depending on the proton fraction $x=n_p/n$; $E_0(n)$ is the energy per nucleon 
in symmetric nuclear matter to which in the case of pure neutron matter 
the asymmetry energy $E_s(n)$ has to be added.
In a recent study of constraints on the high-density behavior of the nuclear 
EoS \cite{Klahn:2006ir}, a set of relativistic mean-field (RMF) EoS of the
Walecka type (NL$\rho$, NL$\rho\delta$) with density dependent coupling
constants and masses (DD, D$^3$C, DD-F, KVR, KVOR) together with that of the
Dirac Brueckner Hartree-Fock (DBHF) approach based on the Bonn-A 
nucleon-nucleon potential has been considered to which we will refer here as a 
hadronic baseline, see Fig. 3. 
Note that all these EoS describe the properties of symmetric nuclear matter at 
the saturation density $n_0\sim 0.16$ fm$^{-3}$ but differ considereably in 
the high-density behavior.
\begin{figure}[t]
\includegraphics[width=0.45\textwidth,height=0.9\textwidth,angle=-90]{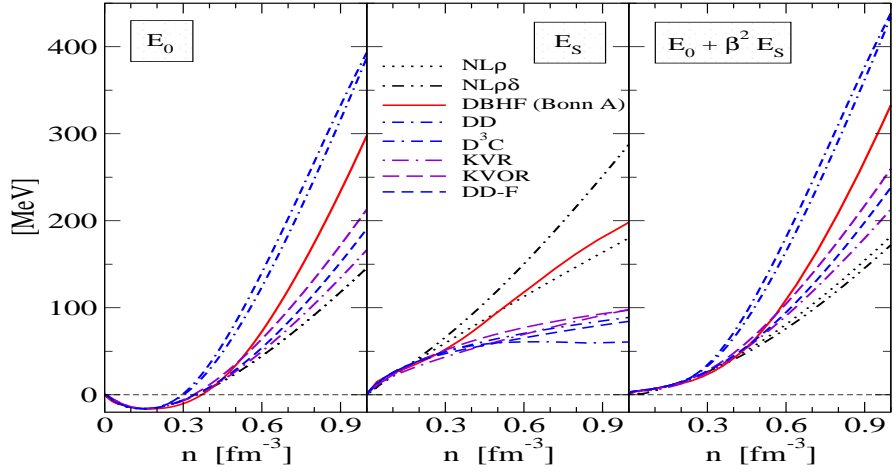}
\caption{Energy per nucleon in symmetric nuclear matter (left), symmetry 
energy (middle) and energy per nucleon in beta-equilibrated neutron star 
matter for the set of relativistic EoS investigated in 
\protect\cite{Klahn:2006ir}.}
\label{fig1a}
\end{figure}
An astrophysical constraint on this behavior comes from the hydrostatic 
equilibrium configurations of spherical stars obtained by solving the 
Tolman-Oppenheimer-Volkoff equations. 
The maximum attainable mass is an important feature of the EoS to be tested 
against observational constraints.
Recently the mass of the pulsar in the double system PSR J0751+1807 has been
determined to $2.1~\pm 0.2~M_\odot$ \cite{Nice:2005fi}, see Fig. 5.
From the analysis of RXTE data on quasi-periodic oscillations in the low-mass 
X-ray binary 4U 1636-536 a measurement of the innermost stable circular orbit 
frequency $\nu_{ISCO}$  corresponding to a mass of 
$1.9-2.1~M_\odot$ \cite{Miller:1996vj} has been reported \cite{Barret:2005wd}, 
see Fig. 4.
These constraints favor a stiff high-density behavior of the EoS as displayed
for DD, D$^3$C and DBHF. 
\begin{figure}[h!]
\includegraphics[width=0.5\textwidth,height=0.75\textwidth,angle=-90]{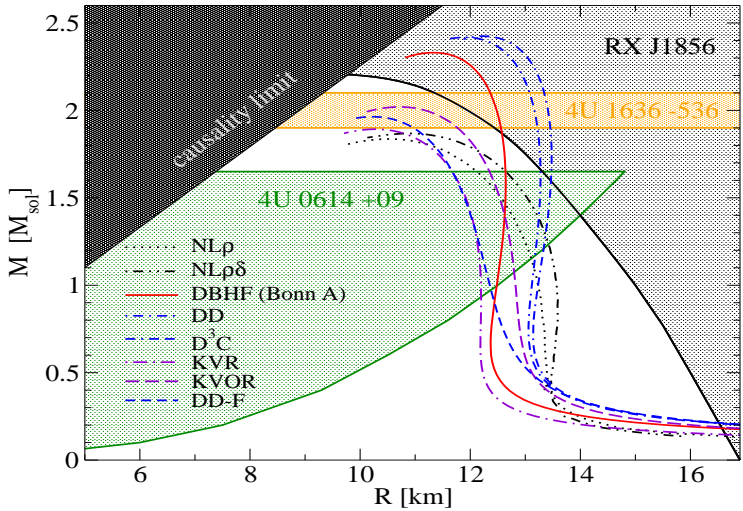}
\caption{Mass-radius constraints from compact stars compared to 
results for a set of relativistic EoS
\protect\cite{Klahn:2006ir}.}
\label{fig1b}
\end{figure}
Note in Fig. 5 that for stiff EoS the maximum masses are at values above
$2.3~M_\odot$ but at a lower central density than for soft EoS which 
allows larger star radii of $R\sim 12-13$ km, see Fig. 4.
The mass-radius constraint obtained from the observation of the thermal 
emission of the isolated neutron star RX J1856.5-3754 seems to favor the stiff
high-density behavior too \cite{Trumper:2003we}, see Fig. 4.
\begin{figure}[h!]
\includegraphics[width=0.5\textwidth,height=0.8\textwidth,angle=-90]{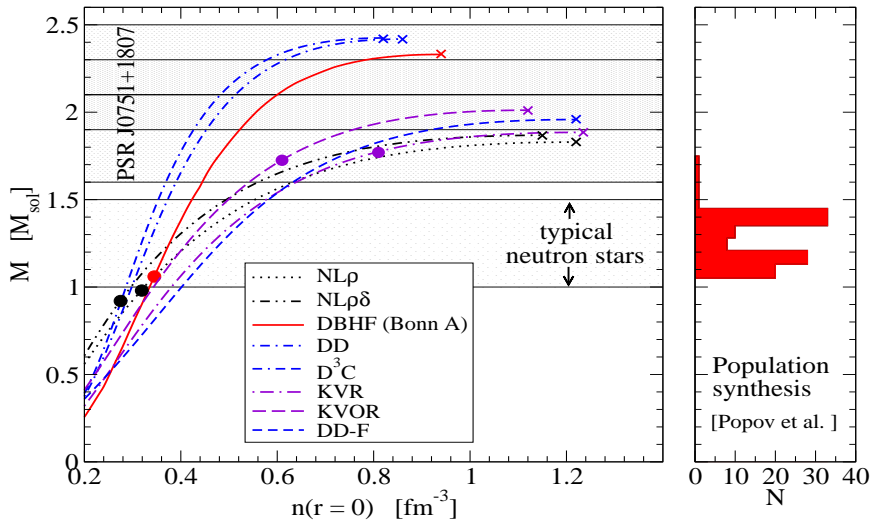}
\caption{Mass constraint from PSR J0751+1807 with 1$\sigma$ and 2$\sigma$ 
error ranges \cite{Nice:2005fi}
compared to results for a set of relativistic EoS
\protect\cite{Klahn:2006ir}. Crosses denote maximum mass configurations
and dots indicate the threshold for the direct Urca process. 
The mass distribution of nearby neutron stars obtained from a population 
synthesis model binned over eight intervals, Ref.~\protect\cite{Popov:2005xa}
defines the range of ``typical neutron stars''.}
\label{fig2a}
\end{figure}

\section{Cooling of Hadronic Stars}

A detailed recent analysis of the regulators of the compact star 
cooling and their in-medium effects (such as neutrino emissivities, 
specific heats and thermal conductivities of different components) 
has revealed the decisive role of the direct Urca (DU) process 
\cite{Blaschke:2004vq}.
Once this process gets initiated at a proton fraction of $\sim 14 \%$
(including muons), the cooling gets dramatically enhanced in disagreement with 
observational data for surface temperature versus age. 
In Fig. 6 the sensitivity to minor changes in the neutron star mass is
demonstrated in such a case.
This problem cannot be solved by a suppression of cooling regulators due to
superfluid pairing gaps since in this case very efficient pair breaking and 
pair formation emissivities will lead to strongly enhanced cooling, again
in disagreement with the data  \cite{Grigorian:2005fn}.
This observation leads to the formulation of the DU constraint:
Hadronic DU processes should not occur for star masses below or within the 
region of typical neutron stars $1.0~M_\odot\le M_{\rm typ}\le 1.5~M_\odot$ 
following from a population synthesis, see Fig. 5, where the DU thresholds are
marked by fat dots.
\begin{figure}[h!]
\includegraphics[width=0.45\textwidth,height=0.75\textwidth,angle=-90]{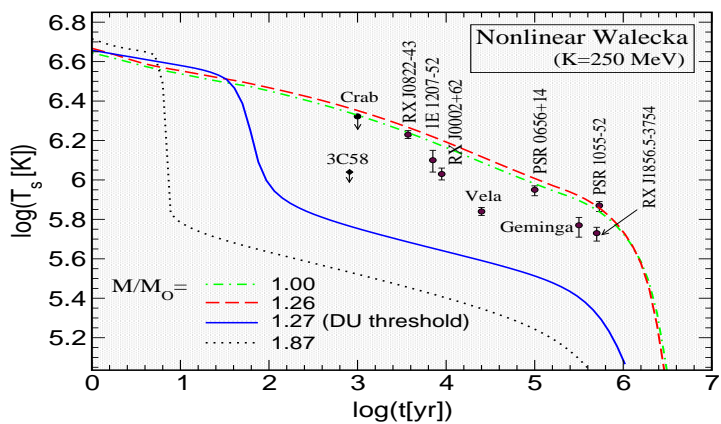}
\caption{Demonstration of the rather unlikely cooling evolution when the DU 
threshold would lie within the interval of typical neutron star masses:
all cooling neutron stars should have the same masses within less than 1\% 
above the DU threshold. Results for standard cooling of a nonlinear Walecka 
model (courtesy: H. Grigorian), see also 
\protect\cite{Blaschke:2004vq,Grigorian:2005fn}.}
\label{fig2c}
\end{figure}
A detailed discussion of the modern astrophysical constraints on the 
behavior of the EoS at high densities \cite{Klahn:2006ir}
shows that none of the purely hadronic EoS from the set presented above 
could fulfill all constraints simultaneously. 
This result motivates the application of this test scheme to hybrid EoS 
with a deconfinement phase transition to quark matter. 
First results for a DBHF-NJL hybrid EoS indicate that indeed the problems with
the DU and flow constraints can be solved this way \cite{Klahn:2006}.   

\section{QCD Phase Diagram and Stability of Hybrid Stars}

For the discussion of the QCD phase diagram in the nonperturbative 
low-temperature/ high-density domain (see Fig.~1) it is customary to use 
effective models of the NJL type \cite{Buballa:2003qv} since asymptotic QCD 
approaches \cite{Schafer:2003jn} are limited to the region of $\mu>500$ MeV 
and Lattice QCD studies have principal problems with the sector $T < \mu$.
DSE studies were still bound to rather schematic interactions 
\cite{Blaschke:1997bj,Roberts:2000aa}, more realistic 
forms of the interaction are at present under consideration and will become
very interesting when the covariant momentum dependence of Lattice QCD studies
of the quark propagator \cite{Bowman:2005vx} could be reproduced
within, e.g., a nonlocal separable quark model 
\cite{Blaschke:2004cc,GomezDumm:2005hy}.
Here we base our report on state-of-the-art results within a NJL model 
for three-flavor quark matter inlcuding a selfconsistent determination of the 
strange quark mass \cite{Blaschke:2005uj}, see Fig. 7. 
Similar results have been obtained by \cite{Ruster:2005jc,Abuki:2005ms}.
\begin{figure}[h]
\includegraphics[width=0.48\textwidth,angle=-90]{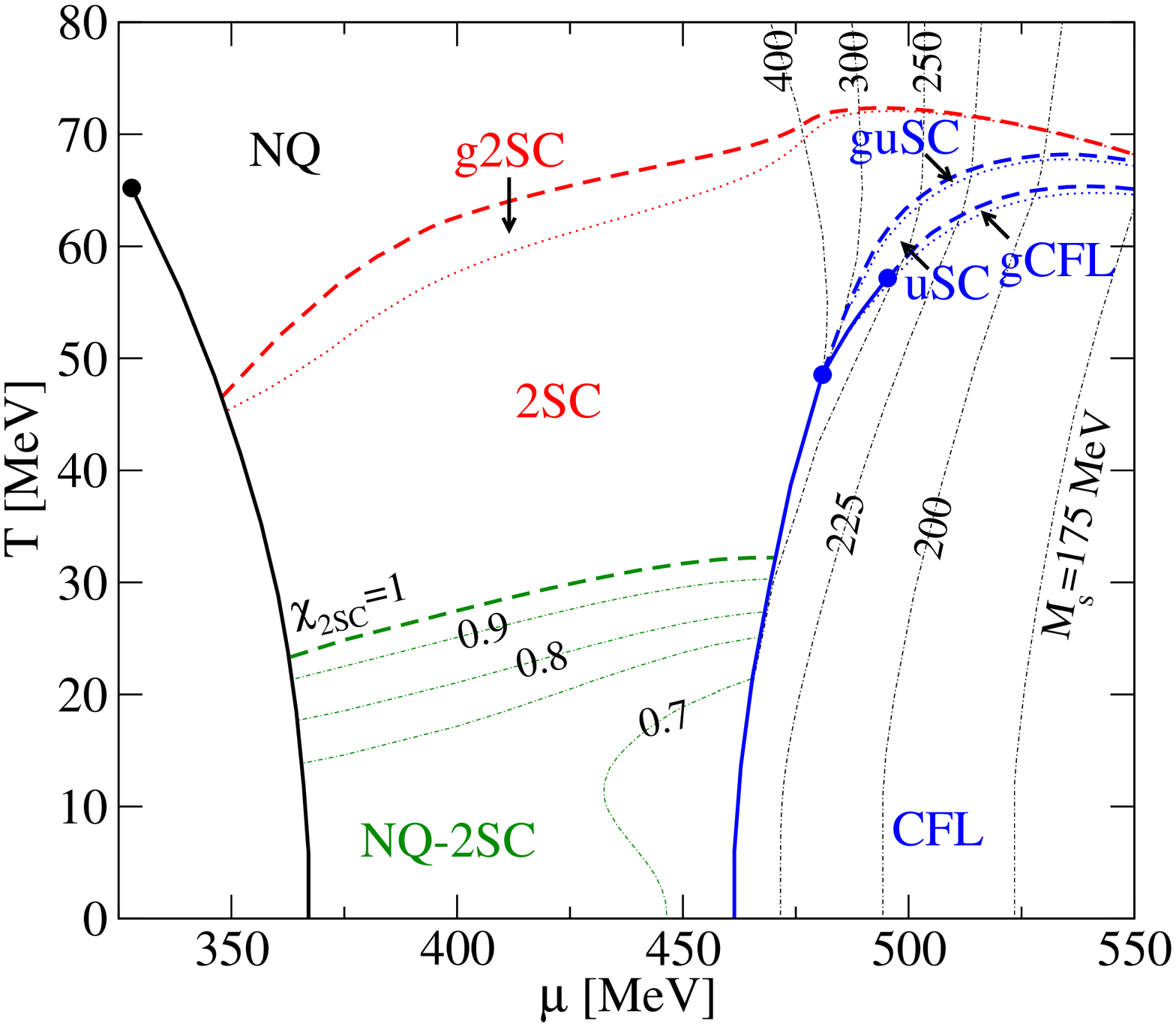}
\hspace{-2cm}
\includegraphics[width=0.48\textwidth,angle=-90]{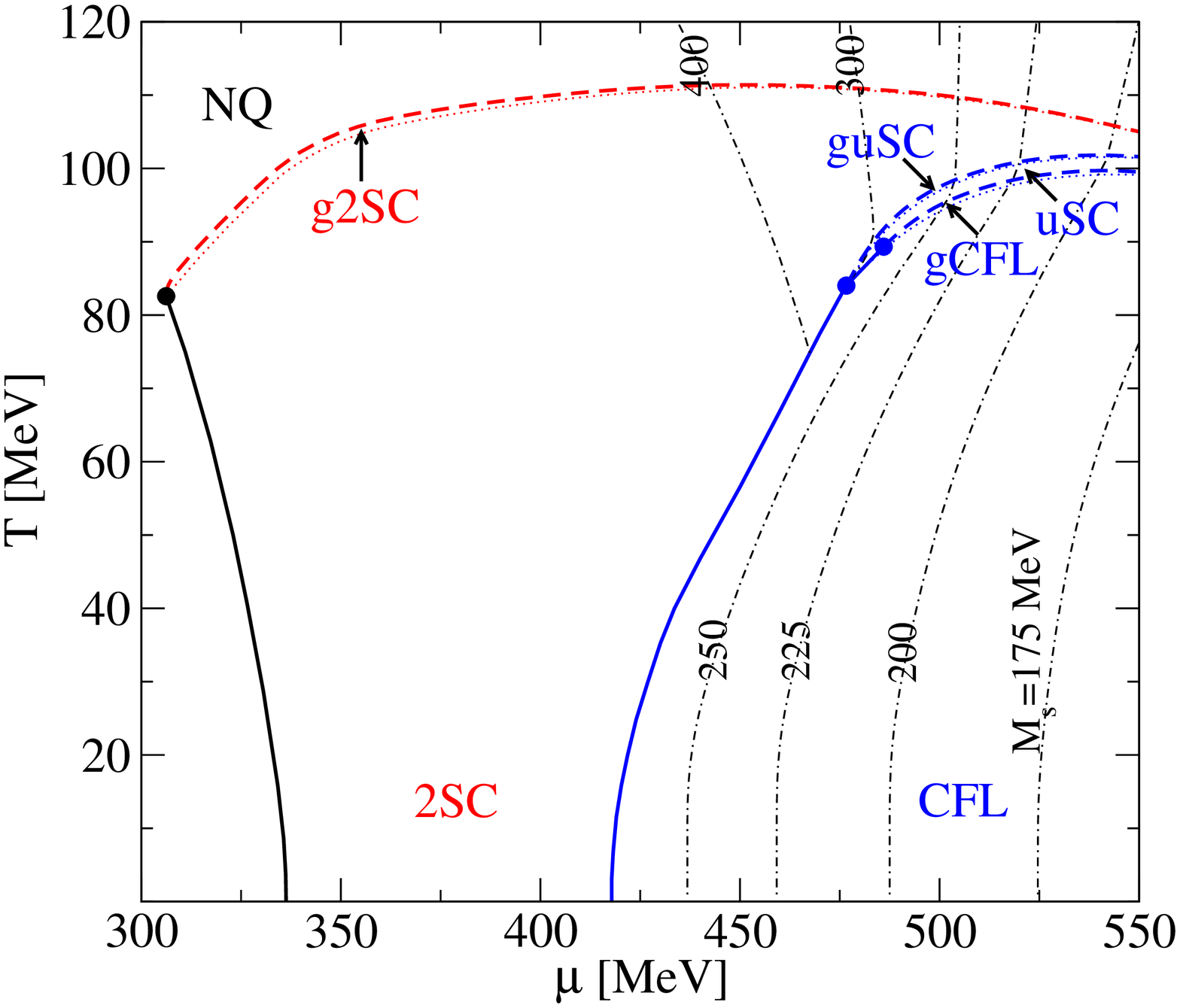}
\caption{Phase diagrams for three-flavor quark matter within the NJL model
for intermediate (left) and strong (right) diquark coupling 
\protect\cite{Blaschke:2005uj}.}
\label{fig3}
\end{figure}
The striking result of these investigations is a sequential melting of the 
light and strange quark condensates due to the large difference
in the dynamically generated light and strange constituent quark masses 
which entails rather different critical chemical potentials for the chiral 
transition in the light and the strange quark sector at low temperatures
and a dominance of two-flavor
superconductivity (2SC phase) in the vicinity of the deconfinement transition.
Three-flavor phases, such as color-flavor-locking (CFL), occur only at rather 
large chemical potentials only when the strange quarks appear and become 
light enough to fulfill the approximate SU(3) flavor symmetry required for 
the CFL phase, see Fig. 7. 
We demonstrate in that Figure that increasing the diquark coupling shifts 
the critical chemical potentials for the onset of color superconducting 
quark matter phases to lower values. 
This leads to an early onset of quark core formation in 
compact star configurations, i.e. to stable hybrid stars already for the 
typical star masses, see Fig. 8.
\begin{figure}[h]
\includegraphics[width=0.45\textwidth,height=0.8\textwidth,angle=-90]
{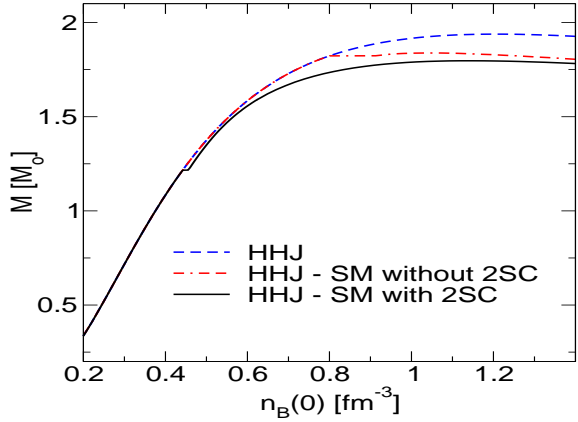}
\caption{Mass-central density relations for hybrid star configurations
with a phase transition between hadronic phase (HHJ: Heiselberg--Hjorth-Jensen
fit to Akmal-Pandharipande-Ravenhall EoS) and quark matter 
(SM: separable model). The occurrence of a two-flavor color
superconducting phase lowers the onset density of the phase transition, see    
Refs.~\protect\cite{Grigorian:2004jq} and \protect\cite{Grigorian:2003vi}.}
\label{fig4}
\end{figure}
Note, however, that once a CFL phase occurs  
the corresponding hybrid star configurations turn out to be unstable against 
collapse \cite{Klahn:2006,Buballa:2003qv}.
Introducing a finite neutrino chemical potential (for the discussion of 
neutrino trapping in the early, hot stages of protoneutron star evolution)  
enlarges the domain of the 2SC in the phase diagram 
\cite{Ruster:2005ib,Klahn:2006}.
Gapless modes in the quark dispersion relations occur when the asymmetry in 
the chemical potentials of the quark species forming the pair exceeds the 
size of the gap, $\delta \mu_{ij}=|\mu_i-\mu_j| > \Delta_{ij}$
\cite{Shovkovy:2003uu}.
Their appearance is therefore tied to the lines of the critical temperatures 
for the vanishing of color superconducting phases in the $T-\mu$ plane where 
gaps become small enough to fulfill this condition.
They  occur here only at high temperatures, not 
relevant for the discussion of (late) compact star evolution, in contrast to
\cite{Alford:2004zr}. 

\section{Hybrid Star Cooling}

For the 2SC phase stable hybrid star configurations with masses even below 
$1.3$ M$_\odot$ have been obtained \cite{Grigorian:2003vi}. 
This phase has one unpaired color of quarks (say blue) for which the 
very effective quark DU process works and leads to a too fast cooling of the 
hybrid star in disagreement with the data.
For details of the cooling calculation see, e.g., Refs. 
\cite{Blaschke:1999qx,Blaschke:2000dy}.
We have suggested to assume a weak pairing channel which 
could lead to a small residual pairing of the hitherto unpaired blue quarks
\cite{Grigorian:2004jq,Blaschke:2005dc}. 
For the resulting gap $\Delta_X$ a density dependence following the ansatz
\begin{equation}
\label{gap}
\Delta_X(\mu)=\Delta_c~\exp[-\alpha(\mu-\mu_c)/\mu_c]
\end{equation}
has been assumed with $\mu_c=330$ MeV being the critical chemical potential.
The choice of parameters $\alpha=25$ and $\Delta_c=5$ MeV has been found to 
give an excellent cooling phenomenology, see left panel of Fig. 9, fulfilling 
a new set of additional constraints \cite{Popov:2005xa}:\\
(i) the brightness constraint \cite{Grigorian:2005fd} given by the
upper barred region in that figure;\\
(ii) the expected number of objects within a mass bin from the population 
synthesis (displayed in Fig. 9  by the darkness of the strip in the T-t plane, 
cf. Fig. 5) is in accordance with actual number of observed coolers.\\
(iii) the Log N - Log S test \cite{Popov:2004ey}, see right panel of Fig. 10.\\
(iv) young coolers like Vela are explained within the mass region of typical 
stars, $M<1.5~M_\odot$.  \\
The physical origin of the X-gap remains to be identified, one possible 
hypothesis is the condensation of color neutral quark sextett complexes
\cite{Barrois:1977xd}. Such calculations have not yet been performed using 
chiral quark models.

For sufficiently small diquark coupling, the 2SC pairing may be inhibited at 
all \cite{Aguilera:2004ag}.
In this case, due to the absence of this competing spin-0 phase with large 
gaps, one may invoke a spin-1 pairing channel in order to avoid the DU problem.
In particular the color-spin-locking (CSL) phase \cite{Schafer:2000tw}
may be in accordance with cooling phenomenology as all quark species are 
paired and the smallest gap channel may have a behavior similar to Eq. 
(\ref{gap}), see \cite{Aguilera:2005tg}. 
A consistent cooling calculation for this phase, however, still requires the 
evaluation of neutrino emissivities and transport coefficients. 
Progress in this direction has recently been obtained 
\cite{Schmitt:2005wg,Jaikumar:2005hy}.

\begin{figure}
\includegraphics[width=0.5\textwidth,height=0.6\textwidth,angle=-90]
{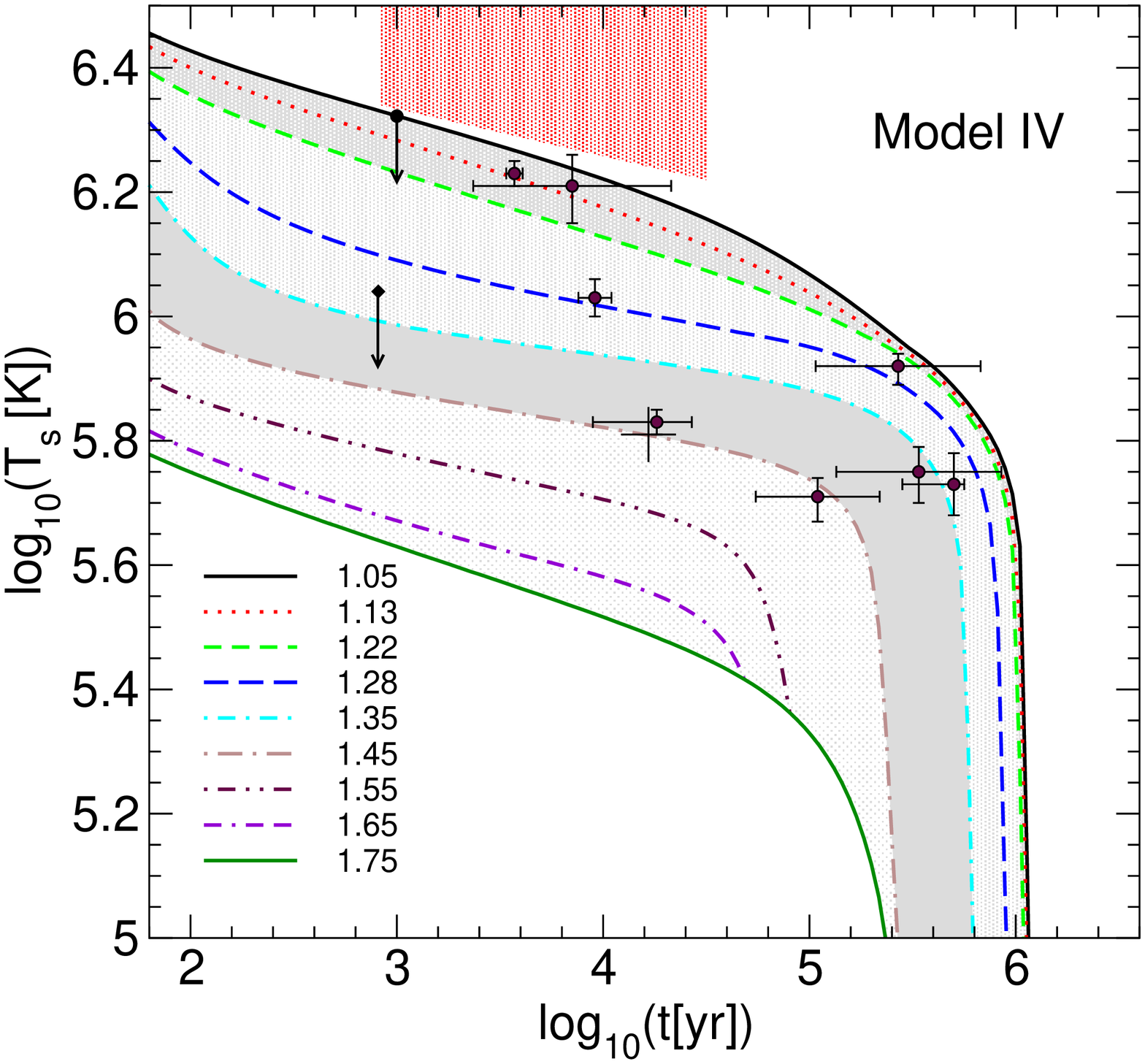}
\hspace{-2cm}
\includegraphics[width=0.5\textwidth,height=0.6\textwidth,angle=-90]
{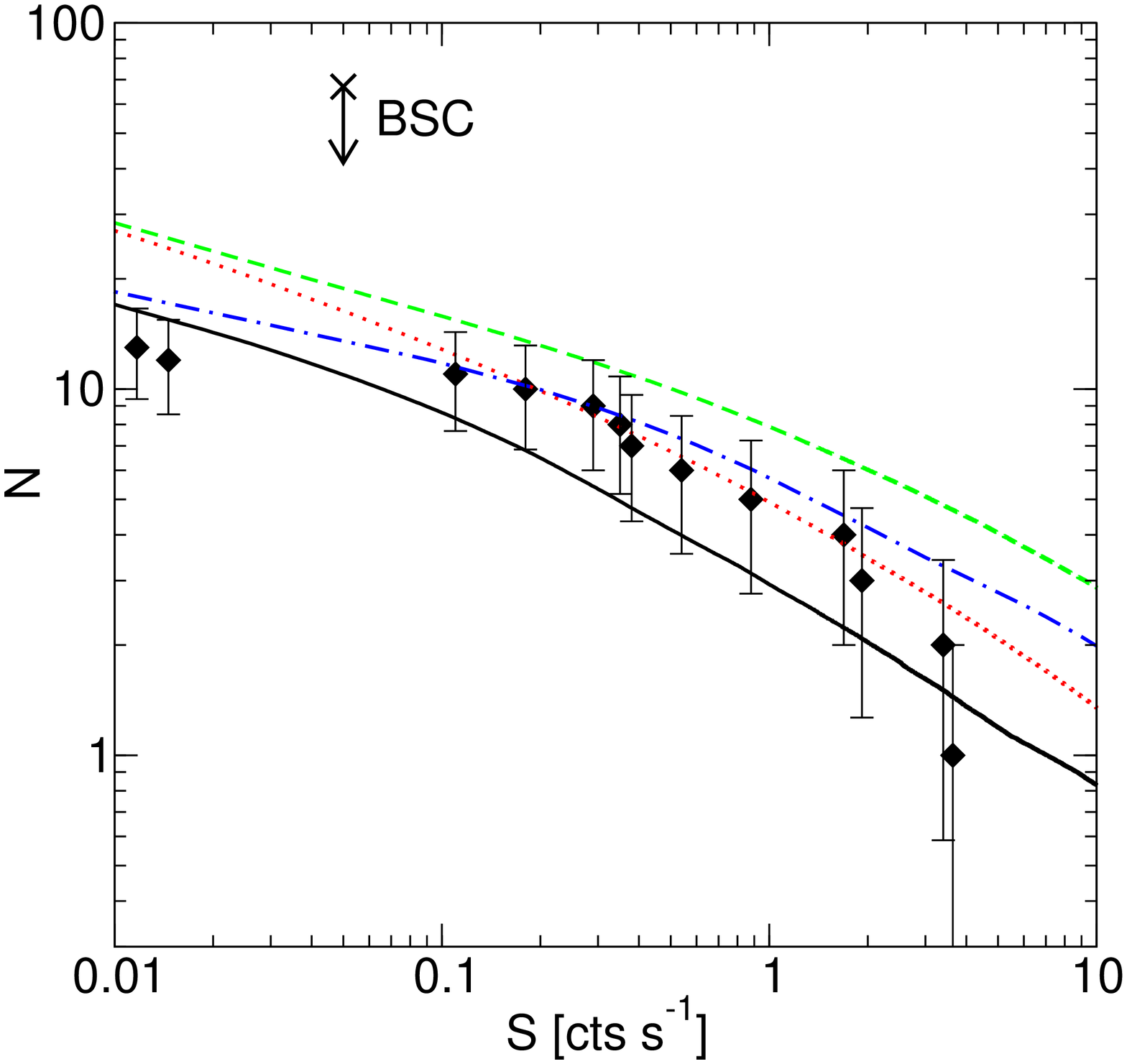}
\caption{Left: Hybrid star cooling curves for model IV of 
Ref.~\protect\cite{Popov:2005xa}.
Different lines correspond to compact star mass values given in the legend
in units of $M_\odot$.
Right: LogN-LogS distribution for the same model }
\label{fig5a}
\end{figure}
\section{Conclusions}
We have shown that  the maximum mass and DU constraints allow for selective 
tests of EoS for QCD matter at high densities. Strange quark matter phases in 
compact stars are not supported by present selfconsistent microscopic 
approaches due to the instability of corresponding configurations.  

Clearly, an improvement of the approaches towards QCD is desired.
One possible path uses the Dyson-Schwinger equation (DSE) approach to the QCD 
partition function \cite{Roberts:2000aa}, from which covariant nonlocal 
models can be derived \cite{Blaschke:2004cc,GomezDumm:2005hy} with 
interaction form factors to be fitted to Lattice QCD.
Our understanding of the confinement/deconfinement  mechanism at finite
temperatures and densities needs to be developed \cite{Blaschke:2000va}.
A promising direction within the QCD DSE approach is a generalization of
the Kugo-Ojima criterion \cite{Alkofer:2003vj}. 
A less ambitious, but effective approach augments the chiral quark dynamics
with the Polyakov loop potential fitted to Lattice data \cite{Ratti:2005jh}.
A general main step is to go beyond the mean field approximation and to
give a consistent description of the hadrons within the favored QCD model. 
First very simple steps within a RMF treatment were presented in Ref. 
\cite{Lawley:2005ru}.
The Mott dissociation of hadronic bound states into the continuum of 
unbound quark states at the chiral/ deconfinement transition will lead to
a proper continuation of hadronic correlations as resonant scattering states
in the quark continuum. 
This concept has been presented long ago within a nonrelativistic potential 
model \cite{Ropke:1986qs} and should be formulated in a proper field 
theoretical manner to address the chiral transition properly. 
This way, an aspect of the interrelation between chiral symmetry restoration 
and deconfinement should be reveale  which need to be worked out.
 
A possible microscopic explanation of the yet unknown X-gap for the 2SC+X 
pairing pattern could come from condensation of bosonic multiquark 
correlations, e.g. quark sextett states in analogy to the quartetting effect 
in nuclear matter \cite{Ropke:1998qs}.
Once baryonic states are consistently included into the description it will 
be clarified whether the diquark coupling (yet considered as a free parameter,
then fixed by the decription of the baryon mass spectrum) will be strong 
enough to allow pairing in the scalar diquark channel. 
The mismatch between up and down quark Fermi levels under $\beta$ equilibrium 
conditions in neutron stars could easily destroy the 2SC phase. 
In this case the spin-1 pairing channels with small gaps, such as the CSL 
phase would turn out as a viable alternative. 
A prerequisite for them to play a role in the explanation of compact star 
cooling would be, however, the absence of the otherwise dominant scalar 
diquark channel(s). For further reading on the fascinating topic of strong 
matter in the heaven we recommend 
\cite{Rajagopal:2000wf,Page:2005fq,Blaschke:2001uj,Blaschke:2006xt}.

\subsection*{Acknowledgements}
I am grateful to all my collaborators who have contributed to the results 
reported in this review with their work,
discussions and support.
Particular thanks go to the collaborations within the Virtual Institute
``Dense hadronic matter and QCD phase transitions'' supported by the Helmholtz 
Association under grant No. VH-VI-041; 
with D.N. Aguilera, J. Berdermann, M. Buballa, A. Faessler, 
B. Friman, C. Fuchs, T. Gaitanos, H. Grigorian, T. Kl\"ahn, D. Rischke,
G. R\"opke, A. Sedrakian, I. Shovkovy, S. Typel, E.N.E. VanDalen, 
D.N. Voskresensky, J. Wambach and H.H. Wolter.
The research would not have been possible without the contributions of 
C.M. Miller, S. Popov, F. Sandin, N.N. Scoccola, R. Turolla,
J. Tr\"umper, F. Weber and V. Yudichev. 
Support by DAAD under grant No. DE/04/27956 and by DFG under grants 
No. 436 ARM 17/4/05 and No. 436 RUS 17/128/03 is acknowledged.

\end{document}